# Noise and spectral stability of deep-UV gas-filled fiber-based supercontinuum sources driven by ultrafast mid-IR pulses

Abubakar I. Adamu,[1, a)] Md. Selim Habib,[2] J. Enrique Antonio Lopez,[2] Peter Uhd Jepsen,[1] Rodrigo Amezcua-Correa,[2] Ole Bang,[1, 3, 4] and Christos Markos[1, 4, b)]

**AFFILIATIONS**

[1]DTU Fotonik, Technical University of Denmark, Kgs. Lyngby, DK-2800, Denmark
[2]CREOL, The College of Optics and Photonics, University of Central Florida, Orlando, FL-32816, USA
[3]NKT Photonics, Blokken 84, Birkerød 3460, Denmark
[4]NORBLIS IVS, Virumgade 35D, DK-2830 Virum, Denmark.
a) Electronic mail: abisa@fotonik.dtu.dk
b) Electronic mail: chmar@fotonik.dtu.dk

**ABSTRACT**

Deep-UV (DUV) supercontinuum (SC) sources based on gas-filled hollow-core fibers constitute perhaps the most viable solution towards ultrafast, compact, and tunable lasers in the UV spectral region. Noise and spectral stability of such broadband sources are key parameters that define their true potential and suitability towards real-world applications. In order to investigate the spectral stability and noise levels in these fiber-based DUV sources, we generate an SC spectrum that extends from 180 nm (through phase-matched dispersive waves - DWs) to 4 µm by pumping an argon-filled hollow-core anti-resonant fiber at a wavelength of 2.45 µm. We characterize the long-term stability of the source over several days and the pulse-to-pulse relative intensity (RIN) noise of the strongest DW at 275 nm. The results indicate no sign of spectral degradation over 110 hours, but the RIN of the DW pulses at 275 nm is found to be as high as 33.3%. Numerical simulations were carried out to investigate the spectral distribution of the RIN and the results confirm the experimental measurements and that the poor noise performance is due to the RIN of the pump laser, which was hitherto not considered in numerical modelling of these sources. The results presented herein provide an important step towards an understanding of the noise mechanism underlying such complex light-gas nonlinear interactions and demonstrate the need for pump laser stabilization.

## I. INTRODUCTION

Fiber-based SC sources are remarkably bright, spatially coherent light sources that can span from DUV to the mid-infrared (mid-IR) spectral region. DUV laser sources, in particular, have numerous important applications in the semiconductor industry[1], such as in photolithography and chip inspection [2], as well as in time-resolved spectroscopy[3]. These applications require a stable and low noise laser source[4]. Although the most stable laser sources are fiber-based [5], many of these fiber lasers use solid-core silica fibers with extremely low loss in the near-IR region, but with extremely high attenuation in the UV and mid-IR regions, rendering them unsuitable for delivery of UV and mid-IR light. Alternatively, solid-core soft-glass fibers, such as ZBLAN and chalcogenide fibers, have been demonstrated to be suitable to provide a spectrum extending into the mid-IR [6,7] and several commercial mid-IR SC sources are now available covering wavelengths up to about 4.9 µm. Single-wavelength mid-IR lasers are now available at around 2 µm (Thulium-doped silica fibers) and 3 µm (Er-doped ZBLAN fibers) [8]. However, solid-core silica fiber based UV laser sources are yet to be realized [9,10]. The main limitations of fused silica for UV sources are, multiphoton absorption [11], radiation-induced photodarkening (also known as solarization) as well as significant material absorption [9].

The fluoride glass ZBLAN has a short wavelength loss edge of about 190 nm and could therefore be used to transmit UV light. However, SC sources require a zero-dispersion wavelength (ZDW) close to the pump, which implies that the core of the ZBLAN fiber must be extremely small to support UV SC generation. Only one demonstration of a UV SC in a ZBLAN fiber has thus been made, in which a unique and never replicated ZBLAN Photonic Crystal Fiber (PCF) was fabricated with a core diameter of about 3 μm. This core diameter was still not small enough to match the dispersion requirements for SC, so the authors had to couple the light to the ~150 nm interstices between holes in the PCF cladding structure to achieve a suitable ZDW that allowed the generation of an SC extending down to 200 nm [12]. The fabrication of ZBLAN PCFs with so small core sizes still remains a challenging task, therefore they are also considered not a viable route towards DUV SC sources.

Hollow-core photonic crystal fibers (HCPF), on the other hand, overcome the limitations of the fiber material, since the light is confined and propagates in a hollow core region, i.e. air. The ability of the HCPF to act as "substrate" and host active and noble gases, has enabled new research directions within the nonlinear fiber-optics field [13,14]. By changing the type of gas and its pressure, both the fiber dispersion and nonlinearity can be tuned [13,14]. Hollow-Core Anti-Resonant Fibers (HC-ARFs) are a sub-category of HCPF defined by broadband transmission and relatively low-loss [15]. These properties, combined with the high laser damage threshold due to a very small overlap of the light with the solid glass material, makes gas-filled HC-ARFs perfect candidates towards ultrafast applications, such as pulse compression[16], multi-octave spanning SC generation [17-19], and tunable DUV sources through resonant DW emission [20]. It has, for example, been shown to efficiently generate high-energy few femtosecond (fs) DW pulses in the DUV and vacuum UV [17,21], which would have a number of important applications [13,14,22]. It should be noted that high energy DW pulses have been also reported using single gas-filled capillaries instead HC-ARFs, but due to their large core size, they require much higher pump pulse energy and peak power than in HC-ARFs [23,24].

A key issue in almost any application of lasers and SC sources is their noise properties. Standard SC sources commercially available use long pump pulses (picosecond or nanosecond) to achieve high average power and have consequently been demonstrated to have high RIN both when pumped in the anomalous dispersion region just above the ZDW (modulational instability or MI based) [25] and in the normal dispersion region just below the ZDW (Raman scattering based) [26]. In other words, MI and Raman scattering are equally noisy processes. The noise of the conventional MI based SC sources is further strongly increased by the subsequent generation of hundreds of solitons that interact in a highly phase and amplitude dependent way. This means two things: First of all the original noise seeding the MI, whether it is quantum or laser technical noise, becomes to some extent irrelevant due to the strong contribution from soliton collisions. Secondly the noise can, to a certain extent, be reduced by special fiber under-tapering to clamp the solitons [27]. Another standard way to strongly reduce the effect of SC noise in applications, such as imaging and spectroscopy, is to use high repetition rates to average out the noise [28].

High repetition rates is generally not an option in gas-filled HCPF-based UV SC sources, because high peak power is required to achieve gas-ionization necessary for DW and SC generation. Except an interesting recent report where MHz-pumped UV SC generation in HC-ARFs demonstrated by compressing an ytterbium fiber laser from 300 fs to 25 fs [29], most of the reports in gas-filled HCPF-based UV SC and DW generation have been using bulky fs lasers with kHz repetition rates. This means that the noise is of fundamental and crucial importance to ascertain the relevance of this new technology for applications.

Unfortunately there has been no experimental report yet on the pulse-to-pulse noise and spectral stability of these sources. Relying upon strong initial self-phase modulation (SPM), which is known to be a coherent effect, several papers have claimed high stability of the SC [13,14,22,30], which has been supported by numerical modelling showing perfect SC coherence[17,31,32]. However, these numerical investigations have neither considered polarization effects nor the noise of the pump laser.

In conventional SPM-based fs-pumped SC generation in solid-core fibers with all-normal dispersion (ANDi), the impact of polarization mode instability (PMI) was demonstrated to be strong in non-polarization maintaining (non-PM) fibers, significantly reducing the pulse and fiber length below which good coherence can be obtained [33]. In most studies of gas-filled HCPF-based UV SC generation, the pump pulses have been shorter than 50 fs, so the PM properties might not play a big role. However, the pump laser noise is critical in any case and its effect on SPM-based fs-pumped SC generation far exceeds the effect of standard quantum noise [34] hitherto used in all numerical noise studies of gas-filled HCPF-based UV SC generation. Furthermore, the fact that the wavelength of the DUV DW generated in gas-filled HCPF-based SC generation is determined by a power-dependent phase-matching condition, implies that any fluctuations of the pump power would directly translate into fluctuations of the wavelength and power of the DUV DW. It was for example already demonstrated that the power of the pump could be used to tune the DUV DW [20], which strongly underlines the importance of a more thorough study of the SC noise, which takes into account the pump laser noise.

Here, we therefore present an experimental and numerical study of the RIN and the long-term stability of gas-filled HC-ARF-based UV SC sources and in particular the RIN of the generated DUV DW and its stability over a duration of 110 hours. We measured the RIN of the strongest DW at 275 nm and compared it with numerical simulations, taking into account both the quantum noise and the actual pump laser fluctuations. It is important to note that although our Ti:sapphire at 800 nm has a measured RIN of 0.2%, the RIN after the Difference Frequency Generator (DFG) was measured to be 5.5% at 2.45μm. Thus, the absolute values of the RIN we measure with this mid-IR pump laser are not representative for Ti:sapphire pumped gas-filled HC-ARF-based UV SC sources, which will be significant lower. However, the general message should apply for any pump laser, i.e., that the noise of such a type of UV SC source, with a UV part directly determined by a phase-matching condition, will be significant and certainly much higher than noise of the pump laser. The simulated results are in good agreement with the measured RIN, clearly underlining the importance of pump laser fluctuations and that these sources are not as coherent as is generally believed from the community. Furthermore, we support our results by an analytical discussion of the influence of the laser fluctuations on the phase matching conditions.

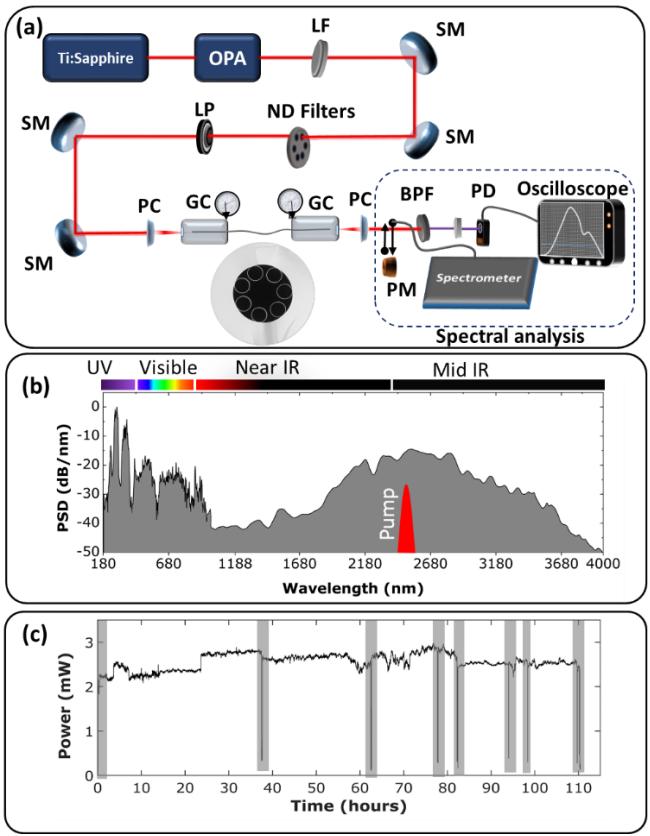

Fig. 1. (a) Experimental setup with a Ti:Sapphire laser pumping an Optical Parametric Amplifier (OPA), long pass filter (LF), silver coated mirror (SM), neutral density filters (ND), linear polarizers (LP) CaFl$_2$ Plano-convex lenses (PC), gas cells (GC), power meter (PM), bandpass filter (BPF), photodiode (PD). Bottom inset: Scanning Electron Microscopy image of the HC-ARF with 44 μm core diameter. (b) Power Spectral Density (PSD) of the generated SC. (c) Total output power versus time over 110 hours. Every dip (shaded in gray) indicates the time of a spectral measurement (change of the beam path from power meter to fiber probe of the spectrum analyzer).

## II. EXPERIMENTAL PART

A single-ring HC-ARF with a 44 μm core diameter and 7 non-touching capillaries (Fig. 1) is filled with argon at 27 bar and pumped in the anomalous dispersion regime at 2.45 μm with ~100 fs (T$_{FWHM}$) and ~8 μJ pulse energy at 1 kHz repetition rate. The experimental set-up used in our experiments is shown in Fig. 1 (a). The pump laser is a standard tunable OPA pumped by an 800 nm Ti:Sapphire laser. The dispersion and loss profile of the HC-ARF used in our experiments can be found in [17], from which it is seen that the pump wavelength is in the anomalous dispersion region inside a low-loss transmission window. An SC spanning from 180 nm to 4 μm is generated, as seen in Fig. 1(b), which has a strong DUV DW at 275 nm followed by a weaker DW at 360 nm.

The average output power of the SC was monitored with a thermal power meter (Thorlabs, C-series) for 110 hours. Minor power fluctuations were observed during the measurement, but without any significant decay, as seen in Fig. 1(c). The DUV spectral profile was recorded at 8 instances over the 110 hours, corresponding to every shaded dip in Fig. 1(c). The spectra are shown separately in Fig. 2(a) and overlaid as gray lines in Fig. 2(b), with the mean spectrum marked by the black curve, which clearly indicates the spectral power fluctuations.

These fluctuations are at the heart of this work and will be characterized in the following in terms of the RIN.

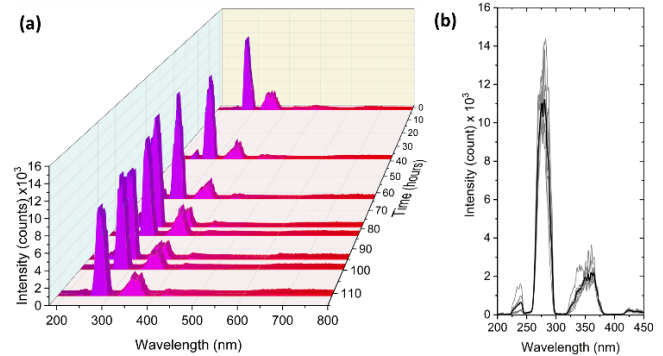

Fig. 2. (a) Long-term stability of DUV DWs measured over 110 hours. (b) Overlay of the measured spectra in grey. The black spectrum signifies the mean of the 8 recorded spectra.

The generated SC from the HC-ARF is collimated and filtered with a 10 nm FWHM bandpass filter with center wavelength at 280 nm (filtered spectra are shown in the supplementary Fig. S2). A narrow bandwidth filter is chosen because SC noise is averaged out with respect to the bandwidth of the filter, i.e., a large bandwidth tends to give lower noise[4]. The filtered SC is then sent to a fast Si detector (Thorlabs DET102, 350 MHz bandwidth, 1 ns rise time). To determine the RIN, a train of 10,000 pulses (or voltage-time series) was recorded with a fast oscilloscope (Teledyne LeCroy - HDO9404 -10 bits resolution, 40 Gs/s, and 4 GHz bandwidth). It is important to note that this setup enables the measurement of pulse-to-pulse intensity fluctuations since the photodiode and oscilloscope are fast enough to detect individual pulses of the 1 kHz SC [35]. The RIN was statistically computed by tracking the peak of every recorded pulse (corresponding to maximum voltage) after the common noise floor level (reference) has been subtracted. The RIN = $\sigma/\mu$ is then defined as the standard deviation ($\sigma$) of the amplitude of the peaks divided by the mean ($\mu$) of the amplitude of the peaks.

The RIN of the main and strongest DUV DW at 275 nm was in this way measured to be 33.3%. The RIN of the secondary but still strong DW at 360 nm was similarly measured using a 10 nm bandwidth filter centered at 360 nm and a fast silicon photodiode (NewFocus, Model 1801, 125 MHz bandwidth). 10,000 pulses were recorded, similar to the measurements performed at 280 nm (filtered spectra in supplementary Fig. S2). The RIN at 360 nm was found to be 8.84 %.

The experimentally observed pulse-to-pulse RIN, measured here for the first time for this type of DUV SC source, contradicts the predictions and numerical conclusions of earlier papers claiming perfect stability [13,14,22] and others presenting numerical modelling of the coherence, showing a perfect coherence of 1[17,20,31,32]. Some de-coherence across the SC was observed in a recent paper, which we will discuss in the last section of the article, but the conclusion was that the UV DW remained largely coherent[20].

The key factor explaining this contradiction is that the earlier modelling of such UV sources did not take into account laser technical noise, i.e., the RIN of the pump laser. Recent numerical and experimental work on SPM-based SC generation with fs pulses in solid-core ANDi fibers [34,36], has however clearly demonstrated that laser technical noise of just 1 % is strongly dominating quantum noise. Our measurements show that the RIN of our pump laser, which is a standard laser for this type of experiments, is 5.5% (see supplementary material Fig. S3).

Physically it therefore makes sense that the noise is significant, since the wavelength of the DUV DW is determined by a phase-matching condition (see supplementary Fig. S1) in which the peak power dependent nonlinear term is very strong due to the high pump peak power, estimated to be $P_0$=75 MW in our experiments. In contrast, the peak powers used in conventional SC sources are in the 10 kW regime, which means that the nonlinear contribution to the phase-mismatch is typically negligible.

## III. THEORY AND EXPERIMENTS OF RIN

To understand the physics behind this noise performance and validate its strength, we simulate the SC generation taking both quantum noise and our measured laser technical noise into account in the initial condition. We use the standard unidirectional pulse propagation equation, which accounts for the plasma effect [32,37,38]:

$$\frac{\partial E}{\partial z} = i\left[\beta(\omega) - \frac{\omega}{v_g} + i\frac{\alpha(\omega)}{2}\right]E + i\left[\frac{\omega^2}{2c^2\varepsilon_0\beta(\omega)}\right]\hat{F}\{P_{NL}\} \quad (1)$$

where $z$ is the propagation distance along the fiber, $t$ is the time in a reference frame moving with the pump group velocity $v_g$, $E=E(z,\omega)$ is the electric field in the frequency domain, $\omega$ is the angular frequency, $\alpha(\omega)$ is the linear propagation loss of the fiber, $c$ is the speed of light in vacuum, $\beta(\omega)$ is the propagation constant, and $\hat{F}\{P_{NL}\}$ represents the Fourier transform of the nonlinear polarization $P_{NL}(z,t)=\varepsilon_0\chi^{(3)}E(z,t)^3+P_{ion}(z,t)$[37,38]. The first term is the Kerr effect, where $\varepsilon_0$ is the vacuum permittivity, and $\chi^{(3)}$ is the third-order nonlinear susceptibility of the noble gas, which here is argon. The second term describes the nonlinear polarization due to molecular or atomic ionization[18,37–39], in which the free electron density was calculated using the quasi-static tunneling ionization approximation and the Ammosov, Delone, and Krainov (ADK) model, described in[40]. Full details of the model and arguments for the used approximations may be found in our earlier paper [17].

A reasonable prediction of the wavelength of the UV DW is by estimating the phase-mismatch $\Delta\beta=\beta_{DW}-\beta_{sol}$ between the propagation constant of the DW ($\beta_{DW}$) and the soliton ($\beta_{sol}$). It should be noted that Eq. (1) does not include an effective modal area and does not have an exact soliton solution. Thus any relation made to Nonlinear Schrödinger (NLS) solitons and the typical nonlinear parameter $\gamma$ is referring to the underlying Generalized Nonlinear Schrödinger (GNLS) type envelope model that does not involve ionization[22]. From the GNLS model the phase-mismatch between the soliton at the pump frequency $\omega_0$ and the DUV DW at the frequency $\omega$ is given by [41,42]:

$$\Delta\beta(\omega) \approx \beta(\omega) - \beta_0 - (\omega-\omega_0)\beta_1 - \beta_{sol} \quad (2)$$

where $\beta_0$ is the propagation constant and $\beta_1 = d\beta/d\omega = 1/v_g$ is the inverse group velocity, both evaluated at the pump frequency $\omega_0$. Several versions of this phase-matching condition for the UV DW generated in gas-filled HCPFs have been proposed using different approximations [20,38]. Here we use the exact $N$-soliton solution to the underlying integrable NLS equation in the GNLS model (obtained by considering only second order dispersion $\beta_2$ and the Kerr effect), which is a bound state between $N$ fundamental solitons with different propagation constants. We match to the one with the largest propagation constant $\beta_{sol} \approx \beta_{sN} = (2N-1)^2/(2L_D)$[43] where $L_D= T_0^2/|\beta_2|$ is the dispersion length and $T_0$ is related to the FWHM as $T_{FWHM}=T_0 \ln(1+\sqrt{2})$. This means that we have neglected the effect from the ionization, which is a good approximation for the UV DW [38]. In the supplementary material we compare all the different versions of the phase-mismatch in terms of validity and their predictions of the UV DW wavelength, showing that Eq. (2) provides the best match to the experiment. The phase-mismatch (plotted in supplementary Fig. S1) predicts a DW wavelength of 238 nm, which is in relatively good agreement with the experimentally measured DW at 275 nm, given the many approximations used (see supplementary Fig. S1).

In our calculation, we included both quantum noise and the measured 5.5% pulse-to-pulse amplitude and pulse width fluctuations from the laser as in [34]. The initial condition with the noise terms becomes:

$$E(0,t) = \sqrt{P_0(1+\Delta_P)} \exp\left[\frac{-t^2}{2[T_0(1+\Delta_T)]^2}\right] + \hat{F}^{-1}\{\Delta_Q\} \quad (3)$$

Here $T_0$ is the pulse duration (60 fs = $T_{FWHM}/\sqrt{4ln2}$), $P_0$ is the peak power (estimated to be 75 MW), and $\hat{F}^{-1}$ is the inverse Fourier transform. The quantum noise $\Delta_Q$ of Eq. (3) is modeled semi-classically as the standard one-photon-per-mode (OPPM) noise added to the initial condition in the Fourier domain as one photon of energy $\hbar\omega_m$ and random phase $\Phi_m$ in each spectral bin $m$ with angular frequency $\omega_m$ and bin size $\Delta\Omega$[44]. The OPPM noise in the frequency domain is given by $\Delta_Q = \sqrt{\hbar\omega_m/\Delta\Omega} \exp(i2\pi\Phi_m)$, where $h$ is Planck's constant and $\Phi_m$ is a random number uniformly distributed in the interval [0,1]. The RIN $\Delta_P$ is Gaussian distributed white noise with zero mean and standard deviation 5.5%. To take into account that our Ti:Sapphire pump laser is a mode-locked laser, we assume that the peak power and pulse length are anti-correlated, i.e., $\Delta_T = -\alpha\Delta_P$, where $\alpha$=1.0 is chosen. Recently an Onefive Origami 10 fs laser was studied, for which $\alpha$=0.4, and anti-correlated amplitude and pulse length noise of only 0.2% was shown to strongly dominate quantum noise [34]. In ref.[36], $\alpha$=1.0 and experimentally measured pump RIN of 1% was used and shown to correctly give the measured noise around the pump.

We numerically calculated the spectral profile of both the RIN and the coherence using 100 spectra from 100 runs with different seeds in the ensemble. When laser technical noise is ignored, the coherence is perfect (see supplementary material Fig. S4), but when the 5.5% RIN is taken into account the coherence is destroyed and the RIN is high, as anticipated (see supplementary material Fig. S5). A direct comparison of the 100 SC spectra overlaid each other clearly shows the significant difference for the two cases (see supplementary material Fig. S6).

In Figs. 3(a) and 3(c) the experimental and numerical average spectra (with both noise sources taken into account) are compared and we see that the numerical model accurately captures the spectral bandwidth and DW at 275 nm, but not the internal DW at 360 nm. Figure 3(b) shows the standard spectral evolution along the fiber, dominated by SPM of the pump and generation of the UV DW once the maximum compression point is reached at ~7.5 cm fiber length.

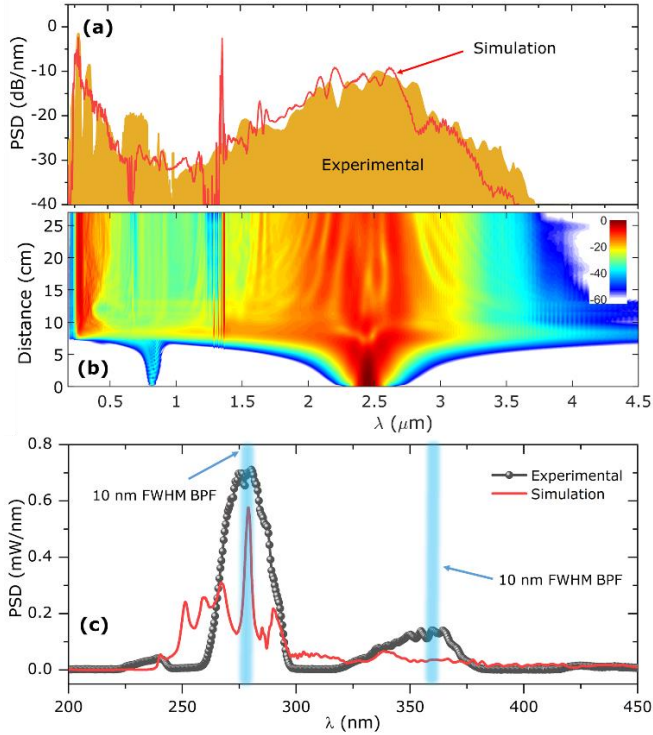

Fig. 3. (a) Numerical simulation and experimental spectrum of broad SC generated in HC-ARF (b) Evolution of spectrum along the length of fiber (c) Numerical and experimental spectra of dispersive waves, plotted in a linear scale. Two filters with 10 nm FWHM (shown in blue) are used to measure RIN at 280 nm and 360 nm.

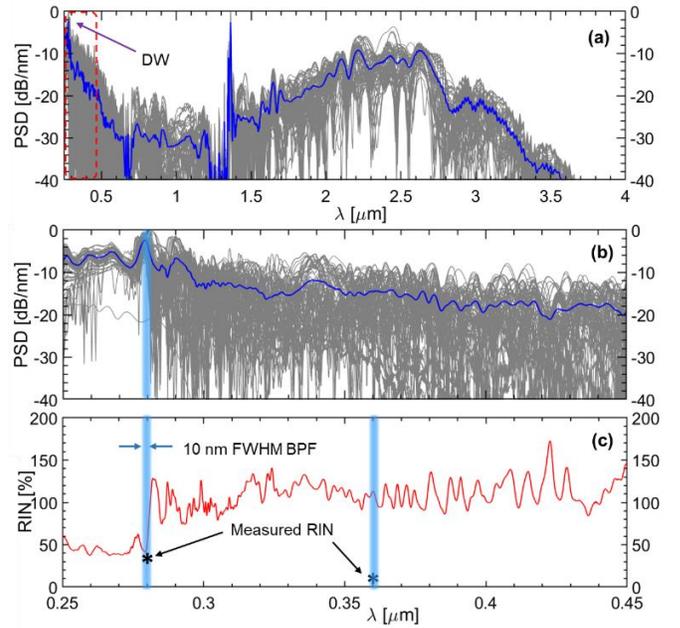

Fig. 4. Numerical simulations: (a) the SC generated by pumping HC-ARF with ~100 fs pulses at 2450 nm and 27 bar Ar pressure. 100 realizations were computed. Blue spectrum shows the average of the realizations. (b) A magnification of the UV section of the spectra. Experimentally the RIN was measured in the blue regions at the DW at 280 nm and 360 nm. (c) Calculated RIN for the 100 realizations plotted in red, with stars indicating the measured 33.3% and 8.84% at 280 nm and 360 nm, respectively. The numerically calculated RIN at 280 nm was found to be 35 %.

In Fig. 4(a, b) we show the numerically calculated average and individual SC spectra with both noise sources taken into account, including a zoom of the UV spectral region. From the pulse-to-pulse statistics, we calculate the RIN shown in Fig. 4(c). In particular we obtain 35% RIN at 280 nm, which matches very well the experimentally found 33.3%. Since negligible noise was found when using only quantum noise (see supplementary Fig. S4), this strongly suggests that laser technical noise is the main reason of the final poor noise performance. Since the internal part of the spectrum at around 360 nm did not perfectly match with our numerical simulations, which is often the case [20,31], the numerically and measured RIN at 360 nm cannot be directly compared.

## IV. DISCUSSION AND CONCLUSIONS

In this work we presented the first experimental study of the long-term stability and pulse-to-pulse noise properties of DUV SC sources based on noble gas-filled HC-ARF fibers. We found that the spectrum and total power fluctuated, but did not show sign of decay over 110 hours of operation. However, our experiments showed that the pulse-to-pulse RIN of the main DUV DW at 275 nm wavelength was 33.3%, which is much higher than predicted or found by numerical modelling in earlier reports on gas-filled DUV SC generation [13,14,17,20,22,30–32].

We have argued from numerical modelling that the observed strong noise originates from the RIN of the pump laser, which was measured to be 5.5%. Our modelling with only standard weak quantum noise (see supplementary fig. S4) confirmed near to perfect coherence (i.e., negligible noise) at all wavelengths, just as in the earlier reports where only this type of noise is considered [14,17,20,31,32]. In contrast we found a RIN of 35% of the shortest DUV DW when taking into account laser technical noise, which is in very good agreement with the experiments. This general result means that the laser technical noise is ultimately limiting the noise performance of SC sources based on coherent fs-pumped SPM, which is in line with recent demonstrations of the noise of SC sources using fs pulses to pump solid-core fibers with all-normal dispersion [33,34,36].

It was found in a recent publication, using numerical modelling with only quantum noise, that the coherence could be not perfect in these DUV SC sources [20]. It is important and very interesting to put the results of this paper into context with our results in terms of soliton numbers and known properties of SC generation. The first key point is that fs-pumped soliton fission based SC generation can be just as noisy as long-pulse pumped MI based SC generation for large soliton numbers $N$, specifically when $N>16$ [45]. The second key point is that in MI-based SC generation any weak noise seed is enough to trigger MI and generate the typically high number of solitons, whose subsequent random interaction will dominate the SC noise. The type and particular strength of the weak seed noise is not important.

In the modelling in [20] 38 fs Gaussian shaped pulses from an $\lambda_0=800$ nm (Ti:Sapphire) laser was used to pump an HCPF fiber with a core diameter of D=44 μm filled with argon at a pressure of 13.5 bar. From Fig. 26 in [14] this gives a nonlinear refractive index of $n_2=1.2 \times 10^{-22}$ m$^2$W$^{-1}$. Assuming that the effective area $A_{eff}$ is the core area, the nonlinear coefficient is then $\gamma = \omega_0 n_2/(cA_{eff}) = 8n_2/(\lambda_0 D^2) = 1.65 \times 10^{-6}$ (Wm)$^{-1}$. From Fig. 1 in [20] we find a group velocity dispersion of $\beta_2 = -2$ fs$^2$/cm, which gives a soliton number of $N=17.9$ and 12.6 for the two pulse energies of 3.0 μJ and 1.5 μJ used in their modelling, respectively. This means that the high pulse energy case in which de-coherence was observed in [20] has a soliton number that was in fact above the threshold of 16 known to lead to highly noisy SC generation for even the very weak quantum noise. The low pulse energy case had a soliton number below the 16 and should thus be highly coherent if only weak quantum noise is considered.

Based on our experimental parameters which are ~100 fs pulses, argon at 27 bar, D=44 μm, and $\lambda_0=2450$ nm, given the $n_2=2.069 \times 10^{-22}$ m$^2$W$^{-1}$ from [14] and $\beta_2=-82.5$ fs$^2$/cm from [17], we find a soliton number of $N=4.66$. Thus we are in the low soliton number case, where complete SC coherence would be expected when using fs-pumped soliton fission based SC generation and taking only the very weak quantum noise into account. Our experiments, confirmed by numerical modelling and analytical considerations of the power dependence of the phase-mismatch between the soliton and DW, demonstrates how pump laser fluctuations, against this expectation, makes the DUV DW and SC have high noise.

Our results clearly reveal the importance of using a low-noise pump laser for DUV SC sources based on gas-filled HC-ARF fibers. Our pump at 2450 nm has a RIN of 5.5% due to how it is generated from the 800 nm Ti:Sapphire seed. However, the mode-locked Ti:Sapphire laser itself, which is the laser most often used to pump these DUV SC sources, has a much lower RIN (in our case it was measured to be 0.28% - see supplementary material S3) and would thus be much more suitable as pump laser. However, a pump laser RIN of 0.28% is still a much stronger noise source than quantum noise for SPM-based SC generation [32], and thus even the Ti:Sapphire pump laser would have to be stabilized to truly enable a future coherent DUV SC source.

## SUPPLEMENTARY MATERIAL

See supplementary material for supporting content.

## ACKNOWLEDGMENT.

This work was supported by Innovation Fund Denmark (6150-00030A & 8090-00060A), Det Frie Forskningsråd (DFF) (8022-00091B), and the Army Research Office (ARO) (W911NF-17-1-0501 and W911NF-12-1-0450) and Air Force Office of Scientific Research (AFOSR) FA9550-15-10041).

The authors would like to thank Shreesha Rao DS and Ivan-Bravo Gonzalo for their helpful discussions and comments during the experiments.

# SUPPLEMENTARY MATERIAL

# Noise and spectral stability of deep-UV gas-filled fiber-based supercontinuum sources driven by ultrafast mid-IR pulses

Abubakar I. Adamu,[1, a)] Md. Selim Habib,[2] J. Enrique Antonio Lopez,[2] Peter Uhd Jepsen,[1] Rodrigo Amezcua-Correa,[2] Ole Bang,[1, 3, 4] and Christos Markos[1, 4, b)]

**ABSTRACT**
This document provides supplementary information to "Noise and spectral stability of deep-UV gas-filled fiber-based supercontinuum sources". Here we provide the details of the phase-matching conditions between the soliton and dispersive waves, and compared our expression to other expressions mentioned in the manuscript. Additionally, we provide he figures for filtered DUV used for the RIN measurements as well as the histograms of the RINs for the pump laser, Ti:sapphire, and DUV at 360 nm and 280 nm. We further compared the coherence and RINs when the pump laser noise is not considered and when considered.

## I. PHASE-MATCHING CONDITION BETWEEN SOLITON AND DW

In the numerical studies of UV SC and DW generation in gas-filled HCPFs the most used model is the unidirectional pulse propagation equation (1) for the full electric field including the higher harmonics, which accounts for the full dispersion, the Kerr nonlinearity, and the plasma effect, but does not include the effective modal area $A_{eff}$ and its dependence on the wavelength, since the transverse profile of the field has simply been ignored. However, in most theoretical considerations of, e.g., soliton and DW interaction, and also in some general modelling [1], the more standard GNLS envelope equation is used, which in the time domain is given by [2]

$$i\frac{\partial A}{\partial z} + i\frac{\alpha}{2}A - \frac{\beta_2}{2}\frac{\partial^2 A}{\partial t^2} - i\frac{\beta_3}{6}\frac{\partial^3 A}{\partial t^3} + \gamma\left(1 + is\frac{\partial}{\partial t}\right)|A|^2 A = 0 \quad (S1)$$

Here self-steepening is present through the parameter $s=1/\omega_0$, where $\omega_0$ is the pump, the Kerr effect is present through the nonlinear parameter $\gamma = n_2\omega_0/(cA_{eff})$, where $n_2$ is the material nonlinearity, and linear fiber loss is present through the parameter α, but the Raman effect is absent, because only noble gasses are considered. Second- and third order dispersion have been included through the parameters $\beta_n = d^n\beta/d\omega^n|_{\omega=\omega_0}$, where $\beta(\omega)$ is the linear propagation constant of the fundamental guided mode. The application of two different models has lead to more than one version of the important phase-matching condition for soliton-DW interaction, which we would like to discuss here. The phase-mismatch $\Delta\beta$ is generally defined as the difference between the propagation constant $\beta_{DW}$ of the DW at frequency $\omega$ and the nonlinear propagation constant $\beta_{NL}$ of the soliton at frequency $\omega_s$. Assuming that the soliton is the pump, $\omega_s = \omega_s$, the phase-mismatch becomes

$$\Delta\beta(\omega) = \beta_{DW}(\omega) - \beta_{NL}(\omega_0) = 0 \quad (S2)$$

In both models people agree on that the DW propagation constant is given by

$$\beta_{DW}(\omega) = \beta(\omega) - \beta_0 - (\omega - \omega_0)\beta_1 \quad (S3)$$

where $\beta_0 = \beta(\omega_0)$ and $\beta_1 = d\beta/d\omega|_{\omega=\omega_0} = 1/v_g(\omega_0)$ is the inverse group velocity, both evaluated at the soliton frequency $\omega_0$. At present, in the literature there exists two different versions of $\beta_{NL}$. If one wants to study DW generation as resonant energy transfer from a soliton, then one needs a soliton solution which is generally found as the fundamental NLS soliton solution to Eq. (1) with $\alpha=\beta_3=s=0$, i.e., ignoring loss and all higher order dispersion and nonlinear effect. The NLS soliton solution has peak power $P_0$, pulse length $T_0$ and propagation constant $\beta_s = \gamma P_0/2$, and is given by

$$A(z,t) = \sqrt{P_0}\,\text{sech}(t/T_0)\exp(i\beta_{s1}z) \quad (S4)$$

Thus, considering phase-matching to the fundamental NLS soliton, one would get $\beta_{NL} = \beta_{s1} = 1/(2L_D)$ and since the dispersion length $L_D = T_0^2/|\beta_2|$ is equal to the nonlinear length $L_{NL} = 1/(\gamma P_0)$ for the fundamental soliton then $\beta_{NL} = \gamma P_0/2$ [3,4]. It is important that this expression is only valid for a fundamental soliton with soliton number N=1, disregarding that the pump pulse typically is stronger and has a higher soliton number N>1, where the soliton number is defined by the relation $N^2 = L_D/L_{NL}$. Travers *et al.* realized that this definition of $\beta_{NL}$ would give too weak a nonlinear contribution to the mismatch and replaced it with $P_C$, defined as the peak power at the point of maximum compression [5,6], which was then estimated to be $P_C = 4.6NP_0$ from numerical modelling based on Eq. (S1)[6]. This gives the following expression for the phase-mismatch

$$\Delta\beta(\omega) = \beta(\omega) - \beta_0 - (\omega - \omega_0)\beta_1 - \frac{\gamma P_C}{2} = 0,$$

$$P_C = 4.6NP_0 \quad (S5)$$

We note that $P_C$ actually has been estimated analytically to be $P_C = NP_0\sqrt{2}$, which has been shown to accurately represent the maximum power of a higher-order soliton for large soliton orders, i.e. the peak power at the point of maximum compression [7]. We have verified this with the N=8 soliton and found $P_C = NP_0\sqrt{2}$ to be very accurate, whereas $P_C = 4.6NP_0$ provides a too high estimate. However, we take a more accurate approach and consider the exact analytical N-soliton solution to the NLS equation [4], which is a bound state of N fundamental 1-solitons with propagation constants $\beta_{sn} = (2n-1)^2\beta_{s1}$, where n=1,2,...,N. The n=N soliton with the largest propagation constant $\beta_{sN} = (2N-1)^2\beta_{s1}$ is also the one with the largest amplitude and smallest pulse length[4], which means that it has the broadest spectrum and is thus the one that spectrally overlaps the most with the DW. We therefore naturally use this propagation constant as $\beta_{NL}$, which gives our phase-mismatch

$$\Delta\beta(\omega) = \beta(\omega) - \beta_0 - (\omega - \omega_0)\beta_1 - \frac{(2N-1)^2|\beta_2|}{2T_0^2} = 0, \quad (S6)$$

The definition of $\beta_{NL}$ as the propagation constant of a soliton means that effects, such as the ionization and self-steepening cannot be taken into account, since no expression for the soliton solutions exists when these higher order terms are taken into account. Using instead a simple plane-wave ansatz to find the nonlinear propagation constant directly from the field equation (1) and then afterwards adjusting the Kerr term to include the nonlinear parameter γ Novoa *et al.* derived the following expression for the phase-mismatch [8]

In the numerical studies of UV SC and DW generation in gas-filled HCPFs the most used model is the unidirectional pulse propagation

$$\Delta\beta(\omega) = \beta(\omega) - \beta_0 - (\omega - \omega_0)\beta_1 - \gamma P_C\frac{\omega}{\omega_0} + \frac{\omega_0\rho}{2n_0c\rho_{cr}}\frac{\omega_0}{\omega} = 0,$$

$$P_C = 4.6NP_0 \quad (S7)$$

where c is the speed of light in vacuum, $n_0$ is the linear refractive index of the gas at the pump wavelength, $\rho$ is the free-electron density, $\rho_{cr}$ is the critical free-electron density at which the plasma is opaque. The Kerr term takes into account shock formation and self-steepening through the factor $\omega/\omega_0$ and does not have the factor ½ since it is not found as a soliton solution, but through plane-wave ansatz. The last term is the ionization term, which also takes into account the factor $\omega/\omega_0$. Novoa et al. showed that the ionization term competes with the Kerr term and allows to explain the generation of mid-IR DWs, but is not very important for the UV DWs [8].

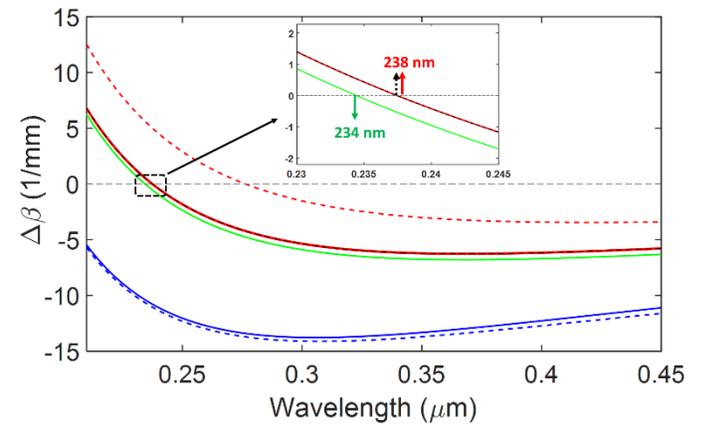

**Fig. S1**: Phase-mismatch curves given as defined in Novoe et.al. [8] with both the Kerr and ionization effects (solid blue) and with only the Kerr effect(dashed blue). The solid green is the phase-match curve described in equation (2) as reported in Mak et. al., The solid red curve is from the phase-matching condition used in this manuscript (equation 2 in the manuscript.) with 2450 nm pump while the dashed red is the same equation but with the dispersive wave matched to 2150 nm pump soliton. The black dashed is when the nonlinear part is ignored.

The experimental parameters and laser specifications are 100 fs pulse width, 2450 nm pump central wavelength and energy of 7.5 x $10^{-6}$ Joules. The effective area of the mode is approximated using [9,10] to be $\approx 1.65\alpha^2$. The dispersion is empirically calculated using [9] for Argon gas in a fiber of 22 μm core radius, cladding radius of 0.68α, with 7 tubes and silica thickness layer of 640 nm. Pressure of 27 bar and temperature of 298 K. The free electron density $\rho$ is 3.87 x $10^{23}$ m$^{-3}$ and critical plasma density, $\rho_{cr}$ = 1.8167 x $10^{26}$ m$^{-3}$, dispersion length 0.437, nonlinearity length 0.021 and $\gamma$ = 6.6447 x $10^{-7}$ W$^{-1}$m$^{-1}$.

## II. SC GENERATION AND NOISE MEASUREMENTS

The experiment was carried out at ambient room temperature of ~23 $^0$C, a custom made gas-cells equipped with CaFl$_2$ windows at one end for passage of incoming coupled light and another for output of SC light before collimation . Prior to taking measurements, the gas cells (interconnected with the hollow core fiber) are purged with 99.99% purity Argon gas to remove any impurity and ambient air from the chamber. To generate the multi-octave SC generation, a 27 bar argon was tuned by manually adjusting the compressed gas regulator and the fixed pressure is maintained throughout the stability measurements in fig. 1c, fig. 2.

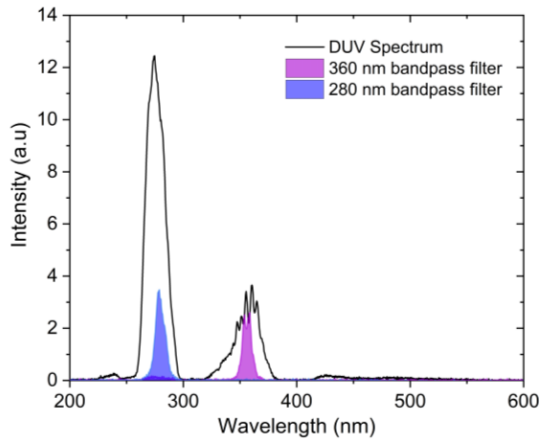

**Fig. S2**. The spectra of filtered DUV dispersive waves. Note: the intensity of the three spectra are not normalized. The FWHM of the filters is 10 nm.

After the broad SC is generated, we filter the 275 nm dispersive waves using a 280 nm central wavelength bandpass filter, with minimum 12% transmission and blocking wavelength range of 200 nm to 10000 nm. Another filter with 360 nm central wavelength is used to filter the relatively low energy DW. The filtered light is shown in purple and blue in Fig. S2. The filtered light is then focused on a fast photodiode with 1ns rise time, connected to an oscilloscope, where a matlab script is used to acquire the peak of each individual pulse. A train of 10,000 pulses was captured for the RIN measurements, and the RIN was computed through the matlab script, histograms shown in fig. S3 with 3 different fits, for all measurements, we take the statistical values for the RIN as some of the distributions are skewed and doesn't necessarily fit a Gaussian, normal or gamma.

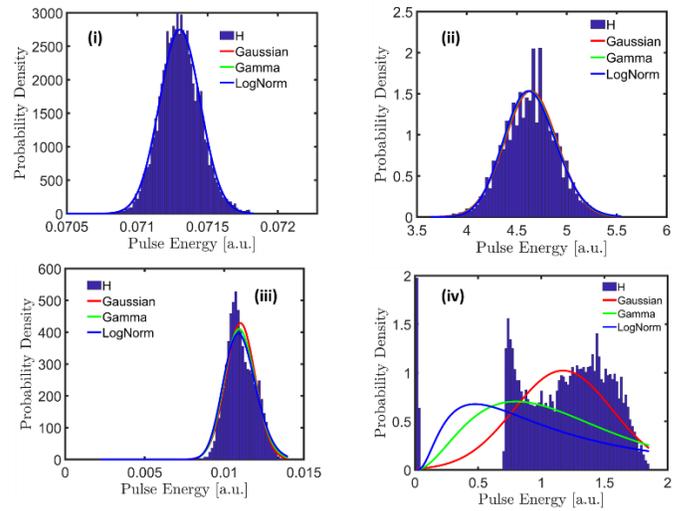

**Fig. S3**. Histograms of filtered pulse energy. (i) Ti:Sapphire at 800 nm with 0.28% RIN (ii) Pump (2.45 μm wavelength) after the TOPAS, with 5.5% measured RIN (iii) 10nm filtered 360 nm UV, with 8.84 % RIN (iv) 10 nm filtered 280 nm DUV with 33.3 % measured RIN. Red, green and blue are gamma, Gaussian and lognormal fits. Note that the RINs used are the statistical values not fits.

## III. CALCULATION OF SC COHERENCE AND RIN

The complex degree of first order coherence of the generated SC was calculated using the following expression[11]:

$$\left|g_{mn}^{(1)}\right| = \left|\frac{\langle A_m^*(\omega)A_n(\omega)\rangle}{\langle|A_m(\omega)|^2\rangle}\right| \quad \textbf{(S6)}$$

The angle brackets in Eq. (S6) represent an ensemble average over the independent simulations. The value of $|g|_{mn}^{(1)}$

indicates quality of the coherence of the SC and primarily a measure of the phase stability[12]. The coherence would be perfect if $|g|_{mn}^{(1)} = 1$, meaning that the electric fields have perfectly equal phase from different laser shots whereas $|g|_{mn}^{(1)} = 0$ indicates the random phase fluctuation from laser shot to shot. The complex degree of first order coherence as a function of the propagation distance is shown in Fig. S4. It can be seen from Fig. S4 that the whole spectrum is fully coherent. However, when we add the pump laser noise of 5.5% (measured from the experiment), the coherence of the generated SC drops drastically which is shown in Fig. S5. From the numerical simulations, it is clear that the coherence of the SC is very sensitive to the pump power fluctuation, and we cannot ignore in order to get a realistic value of the SC.

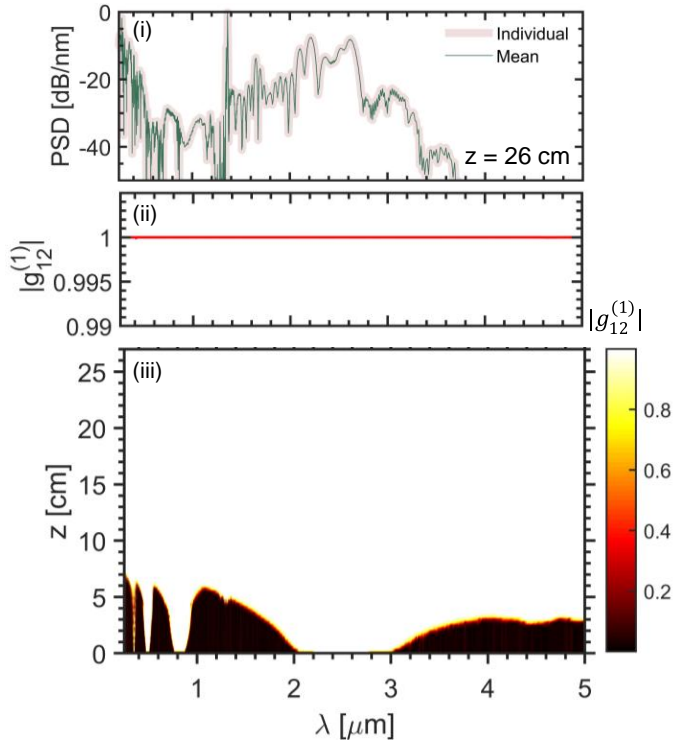
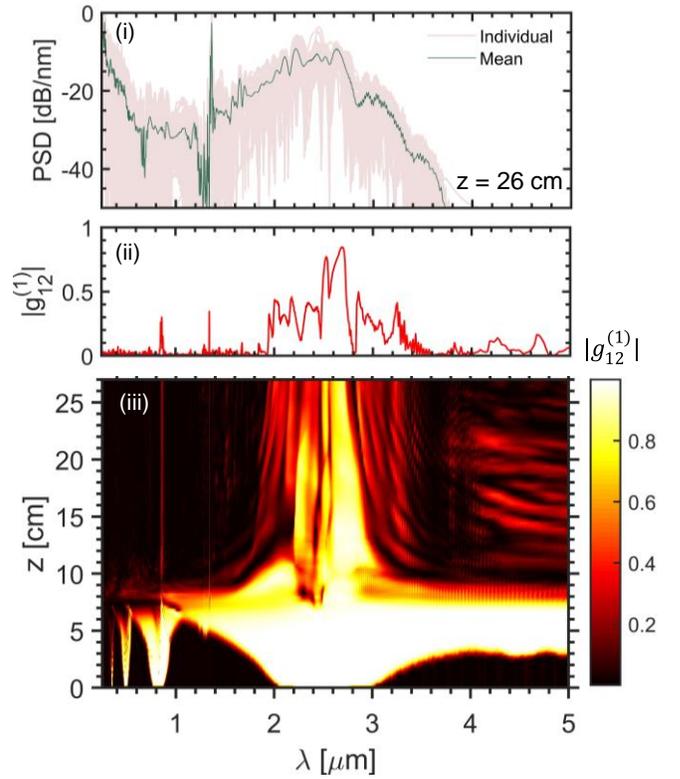

**Fig. S4**. (i) Power spectral density (PSD) at fiber length, z = 26 cm. First order complex degree of coherence for the generated SC, (ii) fiber length, z = 26 cm, (iii) as a function of fiber length. The averaged spectra and coherence properties were found by averaging over 100 simulations with random one photon per mode noise on the optical field, but pump laser noise is ignored.

The relative intensity noise (RIN) was calculated using [3]

$$RIN(\omega) = \frac{<\left(|\widetilde{A_m}(\omega)|^2 - \mu(\omega)\right)^2>^{\frac{1}{2}}}{<|\widetilde{A_m}(\omega)|^2>} \quad (S7)$$

Where μ(ω) is the standard deviation. The calculated RIN is shown in Fig. S6. It can be seen from Fig. S6 that RIN is very low in the full spectrum when pump noise fluctuation is ignored. RIN is increased drastically when pump noise fluctuation in added.

**Fig. S5**. (i) PSD at fiber length, z = 26 cm. First order complex degree of coherence for the generated SC, (ii) fiber length, z = 26 cm, (iii) as a function of fiber length. The averaged spectra and coherence properties were found by averaging over 100 simulations with random one photon per mode noise on the optical field as well as the measured pump laser noise of 5.5% indicating the significant effect of the pump noise to the RIN of the generated spectrum.

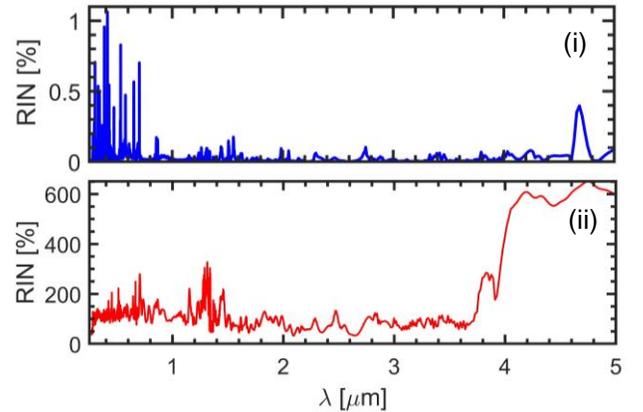

**Fig. S6**. Calculated RIN with 100 realizations for fiber length z = 26 cm (i) without pump noise fluctuations, and (ii) With 5.5% pump laser fluctuations.